\documentclass[12pt,preprint]{aastex}                                          

\begin{document}

\title{A Revised Parallax and its Implications for RX~J185635-3754}
\author{Frederick M.\ Walter and James Lattimer}
\affil{Department of Physics and Astronomy, Stony Brook University,
Stony Brook NY 11794-3800\\fwalter@astro.sunysb.edu, lattimer@astro.sunysb.edu}

\begin{abstract}
 
New astrometric analysis of four WFPC2 images of the isolated neutron
star RX~J185635-3754 show that its distance is 117$\pm$12 pc, nearly
double the originally published distance.
At the
revised distance, the star's age is 5$\times$10$^5$ years, its space
velocity is about 185~km~s$^{-1}$, and its radiation radius inferred
from thermal emission is $R_\infty\sim15$ km, in the range of many
equations of state both with and without exotic matter.  These
measurements remove observational support for an extremely soft
equation of state.  The star's birthplace is still likely to be in the
Upper Sco association, but a connection with $\zeta$~Oph is now
unlikely.

\end{abstract}

\keywords{stars: neutron; stars: individual (RX J185635-3754);
          stars: kinematics; open clusters and associations: Sco-Cen}

\section{Introduction}

The compact object RX~J185635-3754 (Walter, Wolk, \& Neuh\"auser 1996;
Walter \& Matthews 1997) is one of the closest isolated neutron stars
to the Sun \citep{Wal01}.  Because of its relative brightness, its
isolated nature, and its apparently thermal spectrum \citep{PO02}
from X-ray to optical wavelengths, this object affords the
opportunity to study the surface emission properties of neutron stars,
and to measure its radius.  These are important constraints
on the dense matter equation of state and the interior composition of
the neutron star.  The angular diameter has been estimated by
modeling of the spectral energy distribution (e.g., Pons et al.\/ 2002),
while the distance is inferred from the trigonometric parallax.

\cite{Wal01} found a parallax of 16.5$\pm$2.3 milli-arcseconds
(mas), based on three images obtained with the Hubble Space Telescope
Wide Field and Planetary Camera 2 (WFPC2). The implications of the
inferred 60~pc distance include a likely origin in the Upper Scorpius
OB association, possibly as a companion to the runaway O star
$\zeta$~Oph, an age of 0.9$\times$10$^{6}$ years, and a space velocity
of about 100~km~s$^{-1}$. The uniform temperature heavy-element
atmospheric models of \citep{PO02} yield a radiation radius
$R_\infty=R/\sqrt{1-2GM/c^2}\approx6-8$ km and a redshift
$z=(1-2GM/Rc^2)^{-1/2}-1\approx 0.3-0.5$, where $M$ and $R$ are the
neutron star's mass and radius.  The indicated radii and masses are in
the ranges 4.5--8 km and 0.6--1.2 M$_\odot$, respectively.  The
implied radius $R$ is smaller than allowed by any reasonable equation
of state, including that of self-bound quark matter, unless the mass
is less than 0.8 M$_\odot$.  In that case the radius could be matched
by self-bound quark matter configurations.  Relaxing the assumption of
uniform surface temperature, \cite{PO02} showed that somewhat larger
radii, up to R$_\infty$=13~km, are accommodated.
The inferred small radius, together with
the lack of photospheric features in the X-ray spectrum \citep{Bu01}
and the lack of strong pulsations \citep{RGS02}, has led to
speculation that the object might be a self-bound quark star (e.g., Xu
2002; Drake et al.\/ 2002).

\cite{KVA02} reexamined the WFPC2 images, using a more sophisticated
point spread function-fitting technique to measure source positions
and improved (and unpublished) geometric distortion corrections. They
concluded that the parallax is 7$\pm$2~mas, resulting in a distance
about twice what \cite{Wal01} measured.  If true, the observational
support for an extremely soft equation of state is removed.

Three observations are the minimum required to measure a
parallax. Since the expected parallax corresponds to a sub-pixel shift
in the PC camera (45.5~mas pixel$^{-1}$), we sought and were awarded a
fourth observation, which occurred on 2001 March 24, in order to
confirm the parallax.

\section{The Data}

The fourth WFPC2 image was scheduled near the time of
maximum parallactic displacement on March 30.  The first
WFPC2 observation was described by \cite{WM97}, and the next two were
discussed by Walter (2001). The fourth observation is a 7400~second
observation consisting of 6 exposures at two dither positions with
nominal offsets of 5.5 pixels along each axis. The nominal pointing
position was the same as in all of the other observations. The F606W
filter was used to maximize the number of astrometric comparison
stars. The choice of the same filter and pointing position was
designed to minimize differential instrumental distortions.

The particulars of all four observations are summarized in
Table~\ref{tbl-1}. The spectrophotometry has been discussed by \cite{PO02}.

\placetable{tbl-1}

\section{Data Analysis }

We reanalyzed the images obtained at all four epochs, taking into
account corrections discussed by \cite{KVA02}.  We made no attempt to
perform absolute astrometry. All positions are measured with respect
to the first epoch (1996.7).  The images were re-downloaded from the
Multimission Archive at STScI (MAST) prior to analysis to ensure that
the best instrumental calibrations were applied using the MAST
on-the-fly-calibration facility.

We employed two independent measurement techniques and three independent
analysis techniques, as described below.

\subsection{Measurement of Source Positions}

The first measurement of the source positions was performed by
coadding and median-filtering the data obtained at each dither point,
resulting in two images per epoch. We did not analyze the
individual images, primarily because of S/N considerations.  We fitted
the positions of the targets in each pair of images, with a
2-dimensional Lorentzian function as template (using the IDL
MPFIT2DPEAK\footnote{http://astrog.physics.wisc.edu/\~{}craigm/idl/fitting.html
}
function).  We corrected the raw Y positions for
the 34$^{th}$ row error, using the prescription in \cite{AK99}, and
then applied the geometric distortion correction from
\cite{H95}. Uncertainties in the positions are the formal 1$\sigma$
uncertainties of the fit parameters, based on counting statistics in
the images.  To these uncertainties we added in quadrature an uncertainty
of 0.03~pixels to account for systematic effects of non-uniformities in
the intra-pixel response. 
For the subsequent analysis, we used the same reference
objects (mostly field stars) as employed in \cite{Wal01}, except that
stars 109, 121, 125 and 126 were removed because of difficulties in
fitting their positions.

In the second measurement method we used the HSTphot software
\citep{Do00}.  This code fits the point spread functions generated
with the TinyTim \citep{Kr93} software.  We fitted all the images from
each observation, at both dither positions, simultaneously.
\cite{Do00} claims an astrometric accuracy of 0.03 pixels with this
software. The HSTphot astrometry corrects for the 34$^{th}$ row error,
but does not account for the geometric distortions, for which we
applied the \cite{H95} correction. We followed all the processing
steps described in the HSTphot manual prior to running HSTphot.
HSTphot confirms that the objects identified as extended in Table 2
of \cite{Wal01} are indeed extended; all other objects are consistent
with point sources. We did not include the extended objects 104, 108,
or 109 in this analysis.

\subsection{Analysis of the source positions}

We analyzed the measured positions by three independent methods. The derived
proper motions and parallaxes are presented in Table~\ref{tbl-2}.

\subsubsection{Full Astrometric Solution}
First, we performed full $\chi^2$ minimizations for the proper motions
and parallaxes of the objects in the field, including image offsets,
residual rotations from the nominal roll angles, and scale factor
changes from the nominal plate scale (45.5 mas pixel$^{-1}$).
This procedure was performed twice, both excluding and including the
neutron star in the optimization.  In the former case, the proper
motion and parallax of the neutron star were obtained using the image
offsets, residual rotations and scale factor changes from the other
objects.  This distinction was made since the expected proper motion
and parallax of the neutron star are much greater than for those
expected from the field objects.  As anticipated, we found the
parallax and proper motion of the neutron star are slightly smaller
in the second analysis.
The results quoted in Table~\ref{tbl-2} are from the first analysis.

\subsubsection{Independent X,Y Regressions}\label{sec-reg}
Secondly, we determined the proper motion and parallax of the neutron
star independently in the N-S and E-W directions. We registered the
images with the assumption that the mean proper motions and parallaxes
of the field stars are negligible.  We rotated the measured positions
to an equatorial coordinate frame using the nominal roll angles.  We
registered the images by shifting by the median offset in each
coordinate.  We iterated the registration, excluding stars whose
residual differences are significant at $>$3$\sigma$
significance. Registration using a weighted mean shift produced
insignificant differences.  We then determined the deviations from the
nominal roll angle and plate scale by minimizing the diffences between
the positions at each epoch and those of the first epoch.
After resetting the roll angles and the plate scales, we re-registered
the images.  Uncertainties in the image registrations are about 0.02
pixels (1 mas) in each coordinate.

We then shifted the measured position of the neutron star by the plate
offsets, residual rotations and plate scale changes.  The parallax
vector was determined by independent linear regression in both right
ascension and declination. The true parallax is the projection of this
vector in the direction of the parallactic motion (position angle
83$^\circ$).

In both the independent X,Y regressions and the full astrometric
solution we find that the differences from the nominal rotation
($<0.02$ degrees) and plate scale ($<0.03$\%) are small but
significant.  These affect the determination of the parallax at about
the 20\% level.

\subsubsection{Proper Motions From Annual Pairs}

As a safety check, we also determined the proper motions of all
objects using the pairs of observations separated by integral
years. The residuals at the half-year intervals are the sum of
the parallactic shift and the measurement errors.  We found a residual
shifts in right ascension consistent with the parallax of the neutron
star determined in the other analyses, but no significant residual
shift in declination.  The proper motions and parallaxes determined in
this measurement are fully consistent with those in Table~\ref{tbl-2},
but with uncertainties about a factor of three larger.

\placetable{tbl-2}

\section{Discussion and Astrophysical Implications}

The various techniques yield the same results within the
uncertainties, and are presented in Table~\ref{tbl-2}.  For the
subsequent discussion, we use the $117\pm12$ pc distance obtained with
full least squares minimization using the HSTphot positions, which
have slightly smaller nominal errors than those of the 2D fits.  This
result is robust because it is based on 4 observations, two at each
apex of the parallactic ellipse.  This parallax is significantly
smaller than that published earlier \citep{Wal01}. We attribute the
difference to the omission of the geometric distortion of the camera,
which amount to nearly 4 pixels at the edges of the fields
\citep{H95}. The revised parallax is slightly larger than, but agrees
to within errors, with that reported by \cite{KVA02}.

We examined all the stars in the field for their proper motions and parallaxes.
The mean proper motions in (X,Y) are (-1,2)$\times$10$^{-5}$
mas yr$^{-1}$, respectively, and the mean parallax is 4$\times$10$^{-6}$~mas.
This justifies the assumptions used for the image registration in
\S\ref{sec-reg}.



\subsection{Origin and Age}

\citep{Wal01} suggested that RX~J185635-3754 had its origin in the
Upper Scorpius OB association.  This conclusion is unaltered by the
present analysis.  The proper motion is essentially unchanged from
what was originally reported, but the tangential space velocity is
revised to $185 (D/117{\rm~pc})$ km~s$^{-1}$.  For assumed radial velocities
larger than -150 km~s$^{-1}$, the projected path of the neutron star traverses
the projected position of the Upper Scorpius OB association
within the last 2 million years.  

Assuming that the neutron star originated in this association, the 
distance of closest approach of the neutron star to
the center of the association is a function of the assumed radial
velocity. For a present distance of 117~pc, the smallest separation
of the star from the center of the association is about
$8.5\pm2.0$ pc, which occurs for an
assumed radial velocity of -10~km~s$^{-1}$.  The closest approach
occurred in this case about 0.5
million years ago. This separation is appreciably smaller than the
size of the association, extrapolating its present size backwards in
time.

If RX~J185635-3754 was born in the Upper Sco association, and its age
is 5$\times$10$^5$ years, it is no longer a viable candidate for the
binary companion of the runaway O star $\zeta$~Oph. \cite{HBZ01} argue
that the pulsar PSR J1932+1059 is the likely
companion. RX~J185635-3754 is either the result of a more recent
supernova, or is unrelated to the Upper Sco association.

The conclusion of \cite{PO02} that RX~J185635-3754 is, within
uncertainties, on the standard cooling curve expected for neutron
stars is unaltered by the smaller revised age: the luminosity of the
neutron star is also increased.  The estimates quoted in \cite{PO02}
for the magnetic field strength and spin period of the neutron star,
based upon the observation of a bow shock nebula by \cite{vanK00} and
a discussion by \cite{WW02}, are revised to $B\simeq10^{11}$ G and
$P\simeq0.15$ s.
\subsection{The Radius}

The most important reason to measure an accurate distance to this
neutron star is to constrain its mass and radius.  Our measurements
indicate that, to a first approximation, these quantities will be
approximately twice those estimated by \cite{PO02}.  For the
heavy-element model atmosphere fits \citep{PO02} of the spectral energy
distribution of the neutron star, from optical to X-ray wavelengths,
the revised mass and radius of RX~J185635-3754 are shown in
Figure~\ref{fig-1}.  The model atmosphere fits, coupled with the
revised distance of 117 pc, constrain both $R_\infty\simeq15\pm3$ km
and the redshift $z\simeq0.35\pm0.15$.  In turn, the quantities $R$
and $M$ are constrained to $R\simeq11.4\pm2.0$ km and
$M\simeq1.7\pm0.4$ M$_\odot$.  These values are permitted by a large
number of current equations of state \citep{LP01}, including those
containing exotic matter such as quarks in their cores.  However,
these constraints appear to be inconsistent with extremely soft
equations of state.

Since the analysis by \cite{PO02}, new high resolution Chandra X-ray
spectra have become available.  \cite{Bu01} find that the Chandra
X-ray spectra is consistent with a blackbody temperature of 63 eV, and
show that there is no evidence for the absorption lines and edges
expected from a non-magnetized heavy-element atmosphere.  The hotter
temperature requires a smaller angular diameter R$_\infty/D \simeq
0.037$~km~pc$^{-1}$\ for the X-ray emitting blackbody (the statement by Drake
et al. 2002 that the predicted optical flux increases by 10\% is incorrect;
the predicted
optical flux actually decreases by about a factor of 3 relative to 
the extrapolation from the ROSAT blackbody
because the inferred angular diameter decreases by
nearly a factor of 2).  This implies
$R_\infty\simeq4.2$ km, even including a revised, increased distance
\citep{Dr02}.  On this basis, together with the featureless spectrum
and lack of significant pulsations, \cite{Dr02} suggested that
the object may be a self-bound quark star.

However, as demonstrated by \cite{PO02}, and earlier by \cite{Pav96}
and by \cite{RR96}, the inferred radius
depends critically upon the details of the atmosphere and the spectral
energy distribution.  Comparing the ROSAT and EUVE X-ray observations
with optical and ultraviolet data, \cite{PO02} showed that the optical
and ultraviolet radiation cannot originate from the X-ray blackbody
(see Figure~\ref{fig-2}).
Either the X-ray emission arises in a
hot polar cap, or there is significant modification of the spectral
energy distribution from radiative transfer through a stellar
atmosphere.  The ROSAT/EUVE thermal component has an effective
blackbody temperature of 55 eV, so the higher temperature inferred
from Chandra observations only makes this argument more striking.

If the surface has two different thermal components, we can follow the
formalism of \cite{PO02} to estimate the true radius.  Extrapolation
of the 63~eV blackbody fit to the Chandra spectrum
underpredicts the optical flux by
a factor of 6 (Figure~\ref{fig-2}).
Attribution of the optical flux to the cooler part of a
two-temperature blackbody surface results in R$_\infty$/D $<$0.18, or
a maximum radiation radius R$_\infty\sim21$~km at a distance of
117~pc.  In reality, the situation will be more complex than a simple
one or two component blackbody.  Nevertheless, the atmospheric models
of \cite{PO02} have the net effect that the spectral emissions from a
neutron star behave similarly to that of the two component blackbody
model: the total radiating surface area must be substantially larger
than a one component blackbody model to reconcile the optical data and
X-ray data.  On this basis, we suggest that the Chandra data will not
dramatically change the conclusions summarized in Figure~1.

We are examining model atmosphere fits to the Chandra spectrum as well
as to the full multiwavelength spectral energy distribution (work in
progress).  \cite{PO02} discussed only non-magnetic model
atmospheres. New generations of magnetic model atmospheres are
becoming available, and may be able to better reproduce the observed
spectral energy distribution, as well as the lack of significant
spectral features and pulsations.  As we cannot yet select which
atmospheric model best represents the full spectral energy
distribution, the radius inferred from blackbody fits to the X-rays
alone represent no more than lower limits to the true radiation
radius.  

We emphasize in addition that the apparent disagreement between the
Chandra and ROSAT spectral fits is troubling, as is a possible
disagreement between the Chandra and EUVE fluxes where their
responses overlap. \cite{PO02} found that the ROSAT and EUVE fluxes
were mutually consistent, in comparison.  There may be
unresolved calibration issues which affect our ability to determine
the radius at the level needed to distinguish various equations of
state.  Until we understand the full spectral energy distribution, it is
premature to infer that the radius is substantially smaller
than that expected from a normal neutron star.

\acknowledgements This research has been supported by NASA through
grants GO~074080196A and GO 081490197A from the Space Telescope
Science Institute, by LTSA grant NAG57978, and by the USDOE grant
AC02-ER-40317.  We acknowledge stimulating converations with F. \"Ozel
concerning magnetized neutron star atmospheres, and appreciate ongoing
discussions with M. Prakash about the implications for nuclear
theory.

\clearpage
\begin{deluxetable}{rlrrr}
\tablewidth{0pt}
\tablecaption{WFPC2 Observation Log\label{tbl-1}}
\tablehead{
 \colhead{Program} & \colhead{Date} & \colhead{Root} & \colhead{Roll} &
                     \colhead{Duration}\\
  & \colhead{UT} & & \colhead{deg} & \colhead{s}  }
\startdata
6149 & 1996 Oct  6 & U3IM01 & 129.5 & 4800 \\
7408 & 1999 Mar 30 & U51G01 & $-$51.6 & 7200 \\
7408 & 1999 Sep 16 & U51G02 & 124.2& 5191 \\
8567 & 2001 Mar 25 & U62501 & $-$52.78 & 7400 \\
\enddata
\end{deluxetable}

\begin{deluxetable}{rrrrr}
\tablecolumns{5}
\tablewidth{0pt}
\tablecaption{Astrometric Solutions\label{tbl-2}}
\tablehead{
 \colhead{Technique} & \colhead{Proper Motion} & \colhead{Position Angle} & 
                     \colhead{Parallax} & \colhead{Distance}\\
  & \colhead{mas yr$^{-1}$} & \colhead{deg} & \colhead{mas} & \colhead{pc} }
\startdata
\cutinhead{HSTphot positions}
full matrix inversion  & 332.3 $\pm$ 0.4 & 100.45 $\pm$ 0.04 &  8.5 $\pm$ 0.9 & 117 $\pm$ 12\\
independent X,Y regressions & 332.7 $\pm$ 0.5 & 100.35 $\pm$ 0.05 &  8.8 $\pm$ 0.9 & 114 $\pm$ 12\\
\cutinhead{2D fit positions}
full matrix inversion  & 331.0 $\pm$ 0.7 & 100.47 $\pm$ 0.07 &  8.5 $\pm$ 1.5 & 117 $\pm$ 20 \\
independent X,Y regressions & 331.0 $\pm$ 0.6 & 100.28 $\pm$ 0.08 &  9.1 $\pm$ 1.1 & 110 $\pm$ 13\\
\enddata
\end{deluxetable}

\clearpage

\figcaption{ Mass-radius diagrams for the uniform-temperature heavy
element atmosphere models, revised from Figure~17 of \cite{PO02}, for
an assumed distance of 117~pc.  Upper and lower panels are for Fe and
Si-ash compositions, respectively.  Solid and dashed curves are for
equations of state labelled following \cite{LP01}.  The dashed line
labelled ``causality'' is the compactness limit set by requiring
equations of state to be causal.  Dotted lines are contours of fixed
$R_\infty$.  The crosses denote the masses and radii of models which
best fit the optical and X-ray data at the indicated distance, and the
hatched regions surrounding them include the nominal errors indicated
in the constraint relations in equations (4)-(7) of \cite{PO02}, as
well as the nominal error in the distance.
\label{fig-1}}

\figcaption{ The spectral energy distribution of RX~J185635-3754. The 
Chandra LETG spectrum is plotted along with the the EUVE and HST fluxes
from \cite{PO02}. The dashed curve is the best blackbody fit to the Chandra
LETG data from
\cite{Bu01}; it underpredicts the optical fluxes.
The dotted curve is the sum of the X-ray blackbody plus a 15 eV blackbody
with an angular diameter five times larger than the X-ray blackbody. The
two fits are indistinguishable at short wavelengths.
\label{fig-2}}


\begin{figure}
\plotone{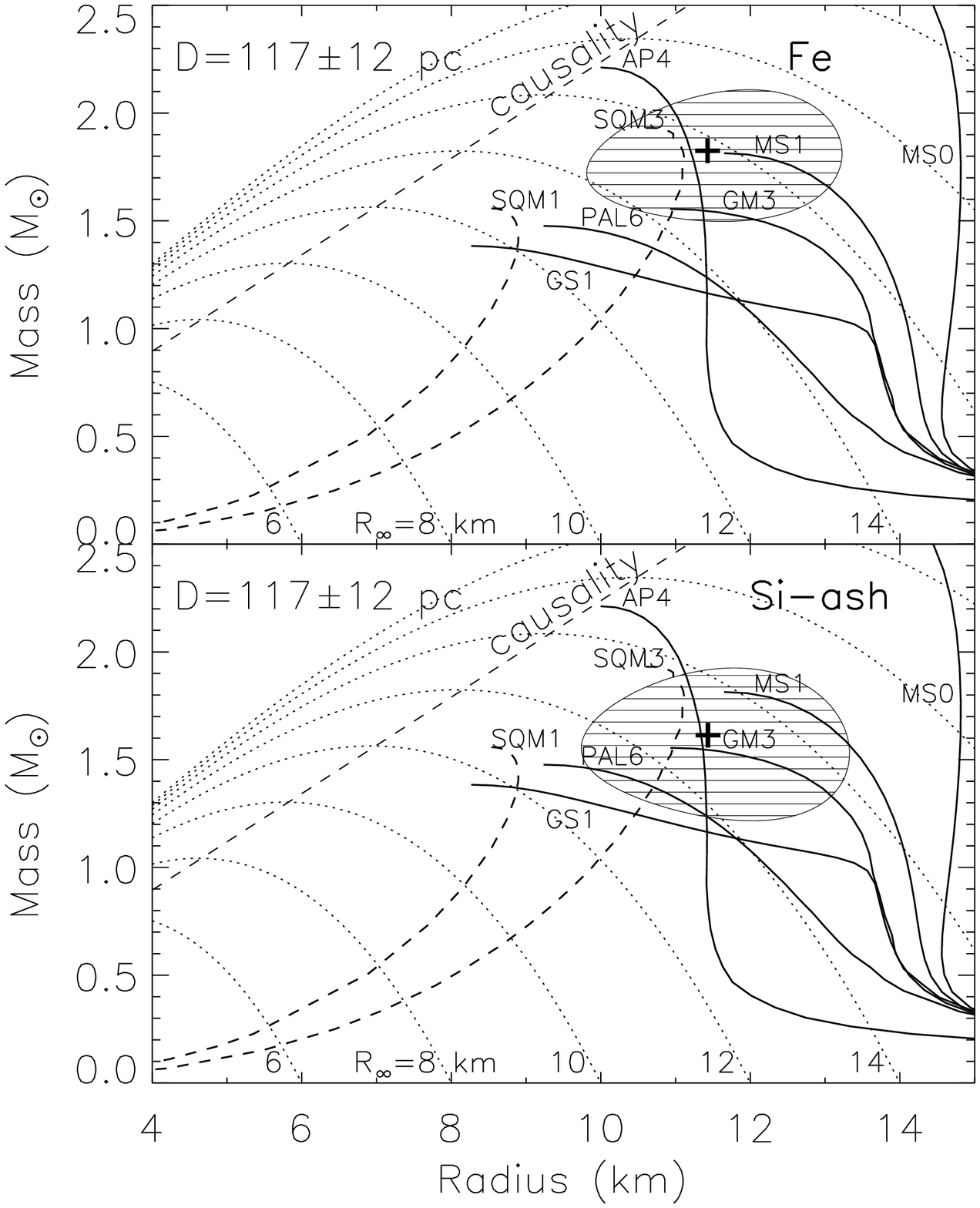}
\end{figure}

\begin{figure}
\plotone{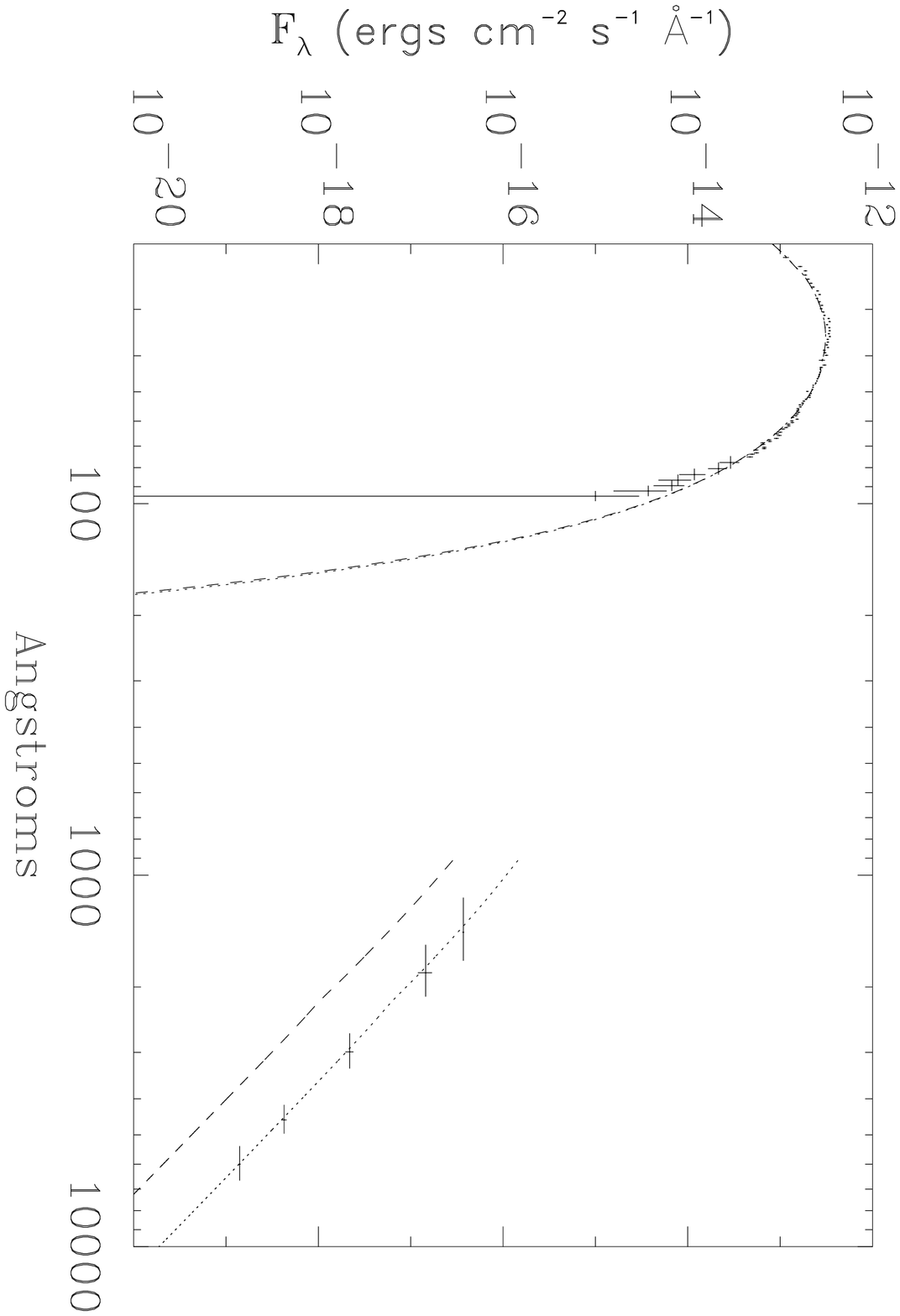}
\end{figure}


\begin{thebibliography}{}

\bibitem[Anderson \& King(1999)]{AK99}
Anderson, J. \& King, I.R. 1999, \pasp, 111, 1095.

\bibitem[Burwitz et al.(2001)]{Bu01}
Burwitz, V., Zavlin, V.E., Neuh\"auser, R. Predehl, P.,
       Tr\"umper, J., \& Brinkman, A.C. 2001, \aap, 379, 35

\bibitem[Dolphin(2002)]{Do00}
Dolphin, A.E. 2000, \pasp, 112, 1383

\bibitem[Drake et al.(2002)]{Dr02}
Drake, J.J., et al.\/ 2002, \apj, in press (astro-ph/0204159)

\bibitem[Holtzmann et al.(1995)]{H95}
Holtzmann, J. et al.\/ 1995, \pasp, 107, 156

\bibitem[Hoogerwoerf, de Bruijne, \& de Zeeuw(2001)]{HBZ01}
Hoogerwoerf, R., de Bruijne, J.H.J., \& de Zeeuw, P.T. 2001, \aap, 365, 49

\bibitem[Kaplan, van Kerkwijk, \& Anderson(2002)]{KVA02}
Kaplan, D.L., van Kerkwijk, M.H., \& Anderson, J. 2002, \apj,
     submitted (astro-ph/0111174)

\bibitem[Krist(1993)]{Kr93}
Krist, J. 1993 in ASP Conf. Ser. 52, Astronomical Data
                  Analysis Software and Systems II, ed. R.J. Hanisch, R.J.V.
                  Brissenden \& J. Barnes (San Francisco: ASP), 536

\bibitem[Lattimer \& Prakash(2001)]{LP01}
Lattimer, J.M. \& Prakash, M., 2001, \apj, 550, 426


\bibitem[Pavlov et al.(1996)]{Pav96}
Pavlov, G.G., Zavlin, V.E., Truemper, J.  \& Neuh\"auser, R. 1996,
   \apj, 472, L33

\bibitem[Pons et al.(2002)]{PO02}
Pons, J.A., Walter, F.M., Lattimer, J.M., Prakash, M., Neuh\"auser, R.,
   \& An, P. 2002, \apj, 564, 981

\bibitem[Rajagopal \& Romani(1996)]{RR96}
Rajagopal, M. \& Romani, R.W. 1996, \apj, 461, 327

\bibitem[Ransom, Gaensler, \& Slane(2002)]{RGS02}
Ransom, S.M., Gaensler, B.M., \& Slane, P.O, 2002, \apjl, in press
   (astro-ph/0111339)


\bibitem[van Kerkwijk \& Kulkarni(2000)]{vanK00}
van Kerkwijk, M. \& Kulkarni, S. 2000, ESO Press Release PR 19/00

\bibitem[Walter(2001)]{Wal01}
Walter, F.M. 2001, \apj, 549, 433

\bibitem[Walter \& Matthews(1997)]{WM97}
Walter, F.M. \& Matthews, L.D. 1997, \nat, 389, 358

\bibitem[Walter \& Wijers(2002)]{WW02}
Walter, F.M. \& Wijers, R.A.M.J. 2002, in preparation

\bibitem[Walter, Wolk \&  Neuh\"auser(1996)]{WWN96}
Walter, F.M., Wolk, S.J., \& Neuh\"auser, R. 1996, \nat, 379, 233

\bibitem[Xu(2002)]{Xu02}
 Xu, R.X. 2002, \apjl, in press (astro-ph/0202365)

\end{thebibliography}
\end{document}